\documentclass[conference]{IEEEtran} %a4paper,10pt,
\IEEEoverridecommandlockouts

\usepackage[a4paper, left=15mm, right=15mm, top=21mm, bottom=44mm,columnsep=0.25in]{geometry}

\usepackage{subfigure,cite} %cite,
\usepackage{amsmath,amssymb,amsfonts,amsthm}
\usepackage{xcolor}
\usepackage{bm,caption}
\usepackage[ruled,vlined]{algorithm2e}
\usepackage{graphicx}%
\usepackage{siunitx,upgreek}
\def\BibTeX{{\rm B\kern-.05em{\sc i\kern-.025em b}\kern-.08em T\kern-.1667em\lower.7ex\hbox{E}\kern-.125emX}}
\DeclareSIUnit{\rad}{rad}

\graphicspath{{figures/}}

\begin{document}

\title{Power-Efficient Deceptive Wireless Beamforming Against Eavesdroppers
\thanks{The work of Yanwei Liu was supported in part by the National Natural Science Foundation of China under grant No. 62371450. The work of Le-Nam Tran was supported  in part by Taighde Éireann - Research Ireland under Grant numbers 22/US/3847 and 17/CDA/4786.}
}

\author{ 
\IEEEauthorblockN{Georgios Chrysanidis\IEEEauthorrefmark{1}, Antonios Argyriou\IEEEauthorrefmark{1}, Le-Nam Tran\IEEEauthorrefmark{2}, Yanming Zhang\IEEEauthorrefmark{3}, and Yanwei Liu\IEEEauthorrefmark{4}}
\IEEEauthorblockA{\IEEEauthorrefmark{1}Department of Electrical and Computer Engineering, University of Thessaly, Greece}
\IEEEauthorblockA{\IEEEauthorrefmark{2}School of Electrical and Electronic Engineering, University College Dublin, Ireland}
\IEEEauthorblockA{\IEEEauthorrefmark{3}Department of Electronic Engineering, The Chinese University of Hong Kong, Shatin NT, Hong Kong,	SAR, China}
\IEEEauthorblockA{\IEEEauthorrefmark{4}Institute of Information Engineering, Chinese Academy of Sciences, Beijing 100093, China\\}
%\IEEEauthorblockA{\IEEEauthorrefmark{5}School of Cyber Security, University of Chinese Academy of Sciences, Beijing, China}
\vspace{-1cm}
}

\maketitle

\begin{abstract}
Eavesdroppers of wireless signals want to infer as much as possible regarding the transmitter (Tx). Popular methods to minimize information leakage to the eavesdropper include covert communication, directional modulation, and beamforming with nulling. In this paper we do not attempt to prevent information leakage to the eavesdropper like the previous methods. Instead we propose to beamform the wireless signal at the Tx in such a way that it incorporates deceptive information. The beamformed orthogonal frequency division multiplexing (OFDM) signal includes a deceptive value for the Doppler (velocity) and range of the Tx. To design the optimal baseband waveform with these characteristics, we define and solve an optimization problem for power-efficient \textit{deceptive wireless beamforming (DWB)}. The relaxed convex Quadratic  Program (QP) is solved using a heuristic algorithm. Our simulation results indicate that our DWB scheme can successfully inject deceptive information with low power consumption, while preserving the shape of the created beam.
\end{abstract}


\section{Introduction}
Eavesdroppers pose one of the most challenging security threats in wireless communication. Their passive operation does not reveal anything regarding their behavior or capabilities. Hence, it is up to the legitimate communication devices to proactively defend against them. The first line of defense against eavesdroppers is covert communication, that is to prevent the detection of the ongoing communication. This approach requires significant modifications to the transmitted waveform to ensure that it has a noise-like form. Another disadvantage of this approach is that it does not prevent localization of the source.

\begin{figure}[t]
	\centering
	\includegraphics[width=0.9\linewidth]{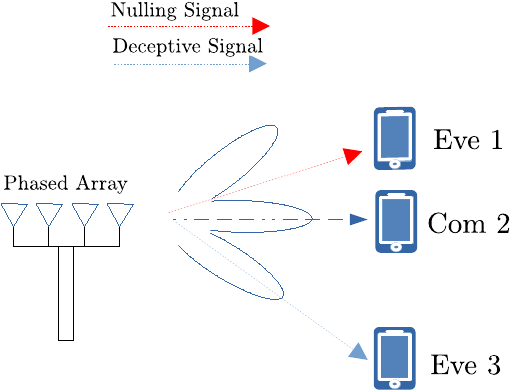}
	\caption{A transmitter has to face two receivers that eavesdrop (Eve 1 \& 3) besides the legitimate receiver (Com 2). With beamforming it can either null the emitted signal to them (Eve 1), or send a deceptive signal with our method (Eve 3).} 
	\label{fig:topology-jcprs}
\end{figure}

In this work we are not interested in addressing the problem of eavesdropping through covertness, but rather in deceiving the eavesdropper. The system model that is used for exploring our idea is illustrated in Fig.~\ref{fig:topology-jcprs} and consists of a single multi-antenna transmitter (Tx) which emits a signal that is intended for a nominal communication receiver (denoted as Com 2). The question we answer is how the transmitter can send data to the intended receiver so as to deceive the eavesdroppers with respect to the transmitter behavior and at the same time minimize the Tx power. In this work we deceive the eavesdroppers by manipulating their perception of their relative distance and Doppler (velocity) to that of the transmitter. We assume wireless communication signals employing the ubiquitous orthogonal frequency division multiplexing (OFDM) scheme. %This is a Multi-Function RF (MFRF) system model shares similarities with~\cite{Tang22} where in that case the JRC system  performed jamming while it was not concerned with covert and deceiving operation. We study different variations of this system.

\textbf{Related Work:} If we do not use complex solutions aimed at covertness, we can minimize the impact of eavesdropping using Tx beamforming. In this way we can null the transmitted signal towards the suspected eavesdroppers~\cite{book:fundamentals-of-radar-signal-processing} (e.g. Eve 1 in Fig.~\ref{fig:topology-jcprs}). Solutions like the one in~\cite{Lagunas20} investigated suboptimal zero-forcing beamforming (ZFB) and nulling so as to improve performance in the low SNR regime when there are more spatial directions that need to be nulled than there are transmit antennas. %This is a problem that has to be addressed in our scenario especially when the number of eavesdroppers is high. %They performed a relaxation of the conventional ZFB problem where they allowed some residual received interference instead of trying to null it out. 
More advanced techniques like injecting artificial noise towards the direction of the eavesdroppers have been explored in~\cite{Liao11}. There is a plethora of systems that aim to control the information emitted to a certain group of receivers (see~\cite{Ahmed19} and references therein). A similar idea is Directional Modulation (DM) which enables secure communication with phased arrays and beamforming~\cite{Daly09}. With DM, beamforming is used to ensure that the signal is distorted outside the intended direction so that incorrect symbols are received. More recent works in Multi-Function RF (MFRF) systems have proposed the joint design of a signal for communications, RADAR, and electronic RF jamming but without considering eavesdropping~\cite{Tang22}. None of the previous works has focused on injecting deceptive information. A first approach to deception focused on compromising the operation of the receiver algorithms at an eavesdropper by injecting frequency variations that purposefully de-synchronize them~\cite{argyriou2020}. This concept of injecting deceptive signals has been shown that it can spoof other parameters of the emitted signal that can be estimated at the receiver such as its Doppler and range~\cite{cnf_2023_radarconf1}. This latest idea has been used for preventing human activity sensing~\cite{jnl_2023_access}, and securing vehicular networks~\cite{chrysanidis2024}. Building on these recent approaches, we extend the idea of spoofing information in OFDM wireless signals in phased array multi-antenna systems capable of Tx beamforming.
		
\textbf{Contributions of this Paper:} In this work we move beyond the idea of nulling the eavesdroppers and instead we propose deceiving them regarding certain location parameters of the emitter. In particular we deceive eavesdroppers with respect to their range and Doppler (velocity) relative to the transmitter that can be passively estimated from the OFDM signals~\cite{Berger10,cnf_2023_radarconf1}. Our scheme does not require any type of out-of-band signaling between the legitimate transmitter and receiver. The reason is that the signal with the artificially generated Doppler and range is the same for every receiver, including the legitimate receivers. However, legitimate receivers only require data demodulation and do not care about Doppler and range. An additional benefit of our scheme is that it deceives eavesdroppers regarding the transmitted information symbols. Hence, our system offers both privacy and security advantages. The previous objectives are achieved by beamforming so that we minimize at the same time the power of the transmitted signal.

%------------LONGER--------------
%\textbf{Paper Organization:} The rest of this paper is organized as follows. Section~\ref{section:system-model} describes the overall signal model at the eavesdropper(s) and the communication receiver(s) as well as our assumptions. Section~\ref{section:eavesdropper-deception} presents the algorithms used at the eavesdroppers for Doppler and range estimation as well as our scheme for deception. Section~\ref{section:optimization} presents the optimization problem we formulate in this work and the solution algorithm. Section~\ref{section:performance-evaluation} presents the performance evaluation of our scheme while Section~\ref{section:conclusions} concludes this paper. 

\textbf{Notation:} Vectors are denoted by bold lower case letters, e.g. $\bm{\theta}$, and matrices with capital bold letters $\mathbf{A}$. The set of complex, $k$-vectors, are denoted by $\mathbb{C}^{k}$, while complex  $m \times n$ matrices belong in the set $\mathbb{C}^{m\times n}$. $\textbf{0}$ denotes the zero matrix, while $\textbf{I}$ denotes the Identity matrix. $[\centerdot]^T$ denotes vector transpose operator. 	

\section{System Model \& Assumptions}
\label{section:system-model} 		
The system model illustrated in Fig.~\ref{fig:topology-jcprs} consists of a phased-array transmitter with $N_T$ antennas that communicates with $N_c$ communications receivers while $N_e$  eavesdropping receivers are present. Each receiver in both groups is equipped with a single antenna. We assume the spatial direction (bearing) of both groups of receivers are known and are located at the angles:
\begin{align*}
	&\bm{\theta_\text{e}}=[\theta_\text{e,1},\theta_\text{e,2},...,\theta_\text{e,$N_e$}]^T,~\bm{\theta_\text{c}}=[\theta_\text{c,1},\theta_\text{c,2},...,\theta_\text{c,$N_c$}]^T.
\end{align*}
The spatial direction can be estimated by transmitting probing waveforms and processing the echo (RADAR functionality) with the same phased array~\cite{Tang22,book:fundamentals-of-radar-signal-processing}. Note that even with estimation errors in the bearing of the eavesdroppers, the emitted signal we will design will still contain the false Doppler and range. 

\subsection{Transmitted Signal Model}
The transmitter emits a baseband signal $\mathbf{S}$$\in$$\mathbb{C}^{N_T\times L}$ from the $N_T$ antennas over $L$ samples to execute \textit{deceptive wireless beamforming (DWB)}. The Tx uses a phased array and in particular a Uniform Linear Array (ULA).

\textbf{Communication Receivers:} The emitted signal $\mathbf{S}$ will be affected by the spatial direction $\theta_{c,i}$ of the $i$-th receiver. In a ULA the steering vector for this communications receiver is:
\begin{align}
	\mathbf{a}_i=[1~e^{j2\pi  \frac{d}{\lambda}\cos(\theta_{c,i})}~...~e^{j2\pi \frac{d}{\lambda} (N_T-1)\cos(\theta_{c,i})}]^T,
	\label{eqn:SV}
\end{align} 
where $\lambda$ is the wavelength and $d$ is the distance between adjacent elements of the array.  We define $\mathbf{A}(\bm{\theta_\text{c}})$ to be the $N_c\times N_T$ steering matrix where the $i$-th row is the steering vector of the $i$-th receiver. If $\mathbf{D}_\text{c}$$\in$$\mathbb{C}^{N_c\times L}$ is the complex time-domain baseband signal we want send to these $N_c$ communication receivers for $L$ samples, then the following must hold:
\begin{align}
	\mathbf{A}(\bm{\theta_\text{c}})\mathbf{S}=\mathbf{D}_\text{c}=[\mathbf{d}_{c,1},\mathbf{d}_{c,2},...,\mathbf{d}_{c,N_c}]^T.
	\label{eqn:AthetaS}
\end{align}
Note that \eqref{eqn:AthetaS} implies a linear constraint since it forces the signal to be transmitted to a specific spatial direction for each receiver, while the information from the other receivers does not interfere towards this direction. In this work $\mathbf{S}$ will be determined through optimization. %With zero-forcing beamforming the emitted signal matrix is $\mathbf{S}$=$(\mathbf{A}^H(\bm{\theta_\text{c}})\mathbf{A}^H(\bm{\theta_\text{c}}))^{-1}\mathbf{A}(\bm{\theta_\text{c}})\mathbf{D}_\text{c}$. 
For each communication receiver vector $\mathbf{d}_c$ will be an OFDM symbol that consists of $L$ samples and has a duration of $T_L$ sec. Hence, $L$ subcarriers are used and placed at frequencies $f_\ell=\ell\Delta f$ Hz with subcarrier spacing $\Delta f=1/T_L$ Hz. If $\mathbf{x}_{c,i}$ is an $L\times 1$ vector that corresponds to the complex QAM symbols transmitted to receiver $i$, then the OFDM symbol of $L$ samples is obtained by applying the IDFT, and thus, \eqref{eqn:AthetaS} becomes:
\begin{align}
	\mathbf{A}(\bm{\theta_\text{c}})\mathbf{S}=\big ( \mathbf{F^{H}}[\mathbf{x}_\text{c,1}~\mathbf{x}_\text{c,2}~...~\mathbf{x}_\text{c,$N_c$}] \big )^T.
	\label{eqn:AthetaS2}
\end{align}
%-----------LONGER
%\begin{align}
%\mathbf{d}_{c,i}=\mathbf{F^{H}}\mathbf{x}_{c,i}
%	\label{eqn:dc}
%\end{align}
$\mathbf{F^{H}}$ is the $L$$\times$$L$ IDFT matrix with time domain data sampled at a rate of $f_s$ Hz: 
\[
[\mathbf{F^{H}}]_{n,\ell}=\frac{1}{\sqrt{L}}e^{j2\pi n\ell/L}=\frac{1}{\sqrt{L}}e^{j2\pi nf_\ell/f_s},0\leq \ell,n\leq L.
\]
Recall that in OFDM the subcarriers $f_\ell$ are placed at orthogonal IDFT frequencies and so the last equation implies $f_\ell=\ell\Delta f =\ell f_s/L$. %The same expression as~\eqref{eqn:dc} holds for the OFDM symbols of the remaining receivers. %Consequently, we can re-write~\eqref{eqn:AthetaS}:
For receiver $i$ we can write the time-domain OFDM symbol as one line of \eqref{eqn:AthetaS2}:
\begin{align} 
\mathbf{S}^T\mathbf{a}_i=\mathbf{d}_i=\mathbf{F}^H\mathbf{x}_i.
\label{eqn:AthetaS3}
\end{align} 

\textbf{Eavesdroppers:} To deceive the eavesdroppers, when we transmit the signal $\mathbf{S}$ towards spatial directions $\bm{\theta_\text{e}}$ we want the combined result of $\mathbf{A}(\bm{\theta_\text{e}})\mathbf{S}$ to correspond to the baseband QAM symbols that are compactly contained in matrix $\mathbf{X}_\text{e}\in \mathbb{C}^{N_e\times L}$. Similar to~\eqref{eqn:AthetaS2} we have:\footnote{The steering matrices can be combined as $\mathbf{A}=[	\mathbf{A}(\bm{\theta_\text{e}})~~	\mathbf{A}(\bm{\theta_\text{c}})]^T$ so that $ \mathbf{A}\mathbf{S}$ is formed.}
\begin{align}
\mathbf{A}(\bm{\theta_\text{e}})\mathbf{S}=\big ( \mathbf{F^H}[\mathbf{x}_\text{e,1}~\mathbf{x}_\text{e,2}~...~\mathbf{x}_\text{e,$N_e$}] \big )^T.
%=\mathbf{D}_\text{e}=[\mathbf{d}_{e,1},\mathbf{d}_{e,2},...,\mathbf{d}_{e,N_e}]^T
\label{eqn:AthetaES}
\end{align}
%$\mathbf{d}_{e,i}$ are the complex samples that should be received at the $i$-th eavesdropper. $\mathbf{A}(\bm{\theta_\text{e}})$ is the $N_e$$\times$$N_T$ transmit beamforming matrix for the eavesdroppers.
	
\subsection{Received Signal Model}
$\mathbf{A}(\bm{\theta_\text{c}})\mathbf{S},\mathbf{A}(\bm{\theta_\text{e}})\mathbf{S}$ represent the emitted signal and its directional transmission due to the phased array, but do not include the impact of the channel. We discuss this part now.

\textbf{OFDM and cyclic prefix (CP) impact:} % Initially we do not distinguish whether the received signal is at a communications receiver or an eavesdropper. %The baseband signal model considers the channel impairments and the steering vector which is unique to each receiver. 
Suppose the LTI channel impulse response $h[q]$ consists of $Q$ taps. The output of the channel is the convolution between its impulse response and the time domain input sequence: $y[l]=\sum_{q=0}^{Q-1} h[q]d[l-q] $. To study this channel let us use the block of OFDM samples $\mathbf{d}_i$ defined in \eqref{eqn:AthetaS3}, that with the addition of the CP, it becomes $\mathbf{d}_i^{'}$. This LTI channel can be written in matrix form as $\mathbf{y}_i=\mathbf{H}_i \mathbf{d}_i^{'}+\mathbf{n}=\mathbf{H}^\text{circ}_i\mathbf{d}_i+\mathbf{n}$ ($\mathbf{H}^\text{circ}_i$ is the circulant channel matrix). To understand the above equation consider the following example with $Q=2$ taps and $L=3$. We use a CP of $Q-1$ samples, which in this example is just sample $d[2]$, and re-arrange $\mathbf{H}_i$:
\begin{align}
\mathbf{H}_i \mathbf{d}_i^{'}&=	\begin{bmatrix}
		h[1] & h[0] & 0 & 0\\
		0 & h[1] & h[0] & 0 \\
		0 & 0 & h[1] & h[0] \\
	\end{bmatrix}\begin{bmatrix}
	d[2]\\	
	d[0]\\
	d[1]\\
	d[2]
\end{bmatrix}\nonumber\\
&=\begin{bmatrix}
 h[0] & 0 & h[1] \\
 h[1] & h[0] & 0 \\
 0 & h[1] & h[0] \\
\end{bmatrix}\begin{bmatrix}	
	d[0]\\
	d[1]\\
	d[2]
\end{bmatrix}=\mathbf{H}^\text{circ}_i\mathbf{d}_i
	\label{eqn:channel_example}
\end{align}
%\AAA{To proceed we first vectorize the matrix of the transmitted signal $\mathbf{S}$, that is $\mathbf{s}=\text{vec}(\mathbf{S})$ with $\mathbf{s}\in \mathbb{C}^{N_T L \times 1}$.} 
%
%
Hence, when $\mathbf{S}$ is transmitted the baseband signal $\mathbf{y}_i\in \mathbb{C}^{L \times 1}$ at a single receiver $i$ is:
\begin{align}
	\mathbf{y}_i=\mathbf{H}^\text{circ}_i\mathbf{d}_i+\mathbf{n}=\mathbf{H}^\text{circ}_i\mathbf{S}^T\mathbf{a}_i+\mathbf{n}.
	%\mathbf{y}_{c,i}	=\underbrace{\alpha_i\mathbf{F^H}\mathbf{P}_i\mathbf{F}\mathbf{A}_i(\theta_{c,i})}_{\mathbf{H}}, \mathbf{s}+\mathbf{n}
	\label{eqn:channel1}
\end{align}
The vector of additive white Gaussian noise (AWGN) samples is $\mathbf{n}$ with a covariance matrix $\mathbf{R_n}=\sigma^2\mathbf{I}$. $\mathbf{H}^\text{circ}_i$ is the channel matrix in the time domain which is circulant because the cyclic prefix (CP) was used. Furthermore, we assume that the channel is static over $L$ consecutive time domain samples (one OFDM symbol) but changes across OFDM symbols due to slow moving receivers. Using~\eqref{eqn:AthetaS3}, we can rewrite~\eqref{eqn:channel1} as:
\begin{align}
	\mathbf{y}_i=\mathbf{H}^\text{circ}_i\mathbf{F}^H\mathbf{x}_i+\mathbf{n}.
	\label{eqn:y}
\end{align}
For each OFDM symbol, the OFDM receiver at the $i$-th eavesdropper first performs the typical OFDM processing. That is it applies a DFT to the $L$ samples of the OFDM symbol in $\mathbf{y}_i$ to obtain the frequency-domain (FD) signal:
\begin{align}
	\tilde{\mathbf{y}}_i=\mathbf{F}\mathbf{y}_i=\mathbf{F}\mathbf{H}^\text{circ}_i\mathbf{F^{H}}\mathbf{x}_i+\tilde{\mathbf{n}}=\mathbf{\tilde{H}}_i\mathbf{x}_i+\tilde{\mathbf{n}}.\label{eqn:y-DFT}
\end{align}
%
%The covariance matrix of the noise component $\mathbf{n}$ is $\mathbf{R_n}$=$\sigma^2\mathbf{I}$. So $\mathbf{R_n}$ is a diagonal matrix (uncorrelated AWGN). %Next we will define the $L\times N_TL$ matrix $\mathbf{H}$.
%	
%
It is known that every circulant matrix $\mathbf{H}^\text{circ}_i$ can be diagonalized using the DFT. That is, $\mathbf{F}\mathbf{H}^\text{circ}_i\mathbf{F^{H}}$ is a diagonal matrix.\footnote{In the general case when time variations introduce significant Doppler within an OFDM symbol and this matrix is not diagonal.} Thus, $\mathbf{\tilde{H}}_i$ becomes a diagonal \textit{frequency domain} channel matrix for receiver $i$ which means that its $\ell$-th diagonal element is multiplied by the $\ell$-th QAM symbol $\mathbf{x}_i[\ell]$. This defines a flat fading channel, i.e., no ISI is present, which is the purpose of OFDM. 

\textbf{Modeling of Flat Fading Coefficients:} Each flat fading channel coefficient in this matrix consists of three elements. First, $\beta_{i}$ is the path loss coefficient. Second, because the $i$-th communication receiver or eavesdropper moves with unknown velocity $v_i$ relative to the Tx, this generates a Doppler shift of $f_{D_i}=v_if_c/c$ \SI{}{Hz} where $f_c$ is the center carrier frequency of the OFDM signal. When $v_i \ll c$ this Doppler effect shift is small. This means that the effect of Doppler in terms of phase shift that it introduces is constant over the duration of an OFDM symbol. In this case we have \textit{slow fading}. We denote this constant phase shift due to Doppler within the $m$-th OFDM symbol as $\exp (j 2 \pi f_{D_i} mT_L)$ (for further discussion regarding this approximation in OFDM RADAR see~\cite{cnf_2023_radarconf1}). Third, if the receiver is located at the unknown distance $R_i$ then the phase shift experienced by the signal on the $\ell$-th subcarrier is
$\exp (-j 2 \pi \ell \Delta f \frac{R_i}{c})$.
Thus, the $\ell$-th diagonal element of this matrix for the $i$-the receiver, and the vectorized version of the matrix are given by:
\begin{align}
[\mathbf{\tilde{H}}_i]_{\ell,\ell}&=\beta_{i,\ell}\exp (-j 2 \pi \ell \Delta f \frac{R_i}{c})\exp (j 2 \pi f_{D_i} mT_L).
\label{eqn:Pi}\\
\mathbf{\tilde{h}}_i(m)&=\text{vec}(\mathbf{\tilde{H}}_i) \nonumber
\end{align}
Note that the vector form $\mathbf{\tilde{h}}_i(m)$ is parametrized according to the OFDM symbol index $m$, and that~\eqref{eqn:y-DFT}, \eqref{eqn:Pi} are useful for the eavesdropper since it wants to estimate $R_i,v_i$. 
%
%The signal emitted towards the $i$-th receiver is that of $\eqref{eqn:dc}$ which when combined with the diagonal matrix $\mathbf{P}_i$ that models the channel, a first version of the signal model for the $i$-th receiver emerges:\footnote{More details regarding the analog model and conversion to digital can be found in~\cite{cnf_2023_radarconf1}.}
%\begin{align}
%\mathbf{y}_i&=\mathbf{F^H}\mathbf{H}_i\mathbf{x}_{c,i}+\mathbf{n}
%\label{eqn:y2}
%\end{align}
%Even though we can write the transmit SNR from this expression, we want to derive a second expression for it as a function of the transmitted signal $\mathbf{S}$ so that the optimization variable is visible. %Since we have vectorized $\mathbf{S}$ we can re-write \eqref{eqn:AthetaS2} for a single communication receiver $i$ at bearing $\theta_{c,i}$ as $\mathbf{A}_i(\theta_{c,i}) \mathbf{s}=\mathbf{F^H}\mathbf{x}_{c,i}$, where $\mathbf{A}_i(\theta_{c,i})\in \mathbb{C}^{L\times N_TL}$ contains $L\times L$ replicated entries of $\mathbf{a}^T(\theta_{c,i})$. 

% which is an expression useful for the Tx.

%\antonis{Consequently, we model the channel from the Tx$\rightarrow$ Receiver with a single complex amplitude plus Doppler. This is ok since we do not assume multiple paths and a frequency selective LTI response.}
	
%\subsection{SNR Expressions}
%To formulate the objectives of our optimization problem we first have to calculate the SNR/SINR when applicable. %Depending on the use of the MFRF system different expressions are applicable.

\textbf{Transmit/Receive SNR:} Regarding the $i$-th communication receiver at bearing $\theta_{\text{c},i}$ its receiver SNR is given by:
\begin{align} 
	\text{SNR}_i=\mathbf{a}^H_i\mathbf{S}^*\mathbf{H}_i^{\text{circ,H}}\mathbf{R}_\mathbf{n}^{-1}\mathbf{H}^\text{circ}_i(\mathbf{S}^T\mathbf{a}_i) =\frac{\|\mathbf{H}^\text{circ}_i(\mathbf{S}^T\mathbf{a}_i)\|^2}{\sigma^2}.\nonumber
%	\label{eqn:sinr-comm-receiver}
\end{align}
We already stated that we do not assume knowledge of $\mathbf{H}_i$ but only the bearing of the receivers in $\mathbf{a}_i$. Thus, the transmit SNR follows the same expression, except without $\mathbf{H}^\text{circ}_i$.
%Hence, the only metric that the transmitter can optimize is the Tx SNR based on the knowledge of the steering matrix:
%\begin{align}\text{SNR}_{\text{Tx},i}=\|\mathbf{S}^T\mathbf{a}_i\|^2/\sigma^2 \label{eqn:sinr-comm-transmitter}
%\end{align} 
%The denominator in the above includes only the AWGN because we do not have interference from the signals towards the other receivers since we use zero forcing beamforming. %If it was different e.g. $\mathbf{A}(\bm{\theta_\text{c}})\mathbf{S}\approx\mathbf{D}_\text{c}$ then \eqref{eqn:channel1} would change and we would have inter-user interference. %(e.g. see ~\cite{Lagunas20} and equations (15),(16) in~\cite{Tang22}).

%\antonis{Check if this ok:} In both cases above we denote $\mathbf{M}=\mathbf{H^H}(\theta_c)\mathbf{R}_\mathbf{n}^{-1}\mathbf{H}(\theta_c)$.

\section{Eavesdropper Deception}
\label{section:eavesdropper-deception}

\subsection{Eavesdropper Speed and Range Estimation}
\label{section:eve-speed-range-estimation}
To understand how eavesdroppers are deceived, we must first understand the algorithms that they use to calculate range and Doppler from OFDM signals. The basic algorithms for passive RADAR functionality from OFDM signals were presented in earlier works~\cite{Berger10,Braun14}, while the scenario of unknown symbols at the eavesdropper was explored in~\cite{cnf_2023_radarconf1}. We now review this concept.

To proceed we first expand the data model in \eqref{eqn:y-DFT} so that it contains data from $M$ successive OFDM symbols, each containing $L$ QAM symbols. $\mathbf{X}_i$ is the $L\times M$ matrix that contains the QAM symbols. Consequently, $\tilde{\mathbf{Y}}$ is the Kronecker product of $\mathbf{X}_i$ and $\mathbf{\tilde{H}^\text{rd}}_i$ with the later containing the range-Doppler information:
\begin{align}
	\tilde{\mathbf{Y}}_i = \mathbf{\tilde{H}}^\text{rd}_i \odot \mathbf{X}_i+\tilde{\mathbf{N}_i},~~
	\mathbf{\tilde{H}}^\text{rd}_i = [\mathbf{\tilde{h}}_i(1),...,\mathbf{\tilde{h}}_i(M)].\nonumber
\end{align}
The next step requires knowledge of $\mathbf{X}_i$. There are two ways to accomplish this. For this part of the algorithm the eavesdropper can use a subset of symbols in a transmitted wireless frame that correspond to known preambles (these are easy to detect sequences at the start of a wireless frame). This ensures accurate knowledge of the preamble $\mathbf{X}_i$. This is a very viable option since wireless communication is carried out with wireless standards that specify preamble structure and contents. This is also the approach used for channel estimation by the standard communication receivers. The second option is to blindly demodulate $\mathbf{X}_i$. This can improve performance as SNR increases as shown in~\cite{cnf_2023_radarconf1}. In either case, after obtaining the symbols $\mathbf{X}_i$, potentially with errors with the second method, the next step is to remove them. We can remove $\mathbf{X}_i$ by performing elementwise division on $\tilde{\mathbf{Y}}_i$:
\begin{align}
	\tilde{\mathbf{Z}}_i=\tilde{\mathbf{Y}}_i \oslash \mathbf{X}_i=\mathbf{\tilde{H}}^\text{rd}_i+\tilde{\mathbf{N}} \oslash \mathbf{X}_i.  \label{eqn:Z_tilde}
\end{align}
%where $\mathbf{N}_f=$.
The OFDM-based passive RADAR algorithm at the eavesdropper performs 2D DFT on the $L \times M$ matrix $	\tilde{\mathbf{Z}}_i$ across the $\ell,m$ indexes with sampling periods $\Delta f$ and $T_L$ (essentially sampling \eqref{eqn:Pi}) respectively, to obtain the range-Doppler response contained in $\mathbf{\tilde{H}}^\text{rd}_i$. From \eqref{eqn:Pi} we see that the peak in this range-Doppler response will occur at positions $R_i/c$ and $f_{D_i}$. With sufficient range and Doppler resolution, which for OFDM is $1/(L \Delta f)$ and $1/(MT_L)$=$\Delta f/M$ respectively, the eavesdropper can estimate its relative speed to the transmitter from the range-Doppler response (range-Doppler images that are generated with this method can be seen in Fig.~\cite{cnf_2023_radarconf1}). 
%
%----------LONGER PAPER----------
% (example plot shown in Fig.~\ref{fig:radar-ofdm-range-doppler1}).\footnote{The reader is referred to~\cite{Braun14,cnf_2023_radarconf1} for more details regarding the range-Doppler response derivation in OFDM RADAR.}
%\begin{figure}[t]
%	\centering
%		\includegraphics[width=0.98\linewidth]{radar-ofdm-range-doppler_SNR10dB_new}
%%\caption{	\subfigure[]}	
%%	\hspace{-0.18cm}
%%	\subfigure[URx SNR of \SI{30}{\dB}. SNR increase results in better demodulation performance for $\hat{X}$ and improved quality signature.]{\includegraphics[width=0.48\linewidth]{../../../bitbucket/ssp/results/paper-ofdm-radar-spoofing/radar-ofdm-range-doppler_SNR40dB_new}}
%	\caption{Range-Doppler responses for different SNR at Eve with a deceptive range at \SI{30}{\meter} and Doppler at \SI{2}{\kilo\hertz}.}
%	\label{fig:radar-ofdm-range-doppler1}
%\end{figure}

\begin{figure*}[!htb]
	\centering
	\subfigure[Nulling with $\theta_c$=$80^o$, $\theta_\text{eav}$=\{$30^o$,$50^o$\}.]{\includegraphics[width=0.329\linewidth]{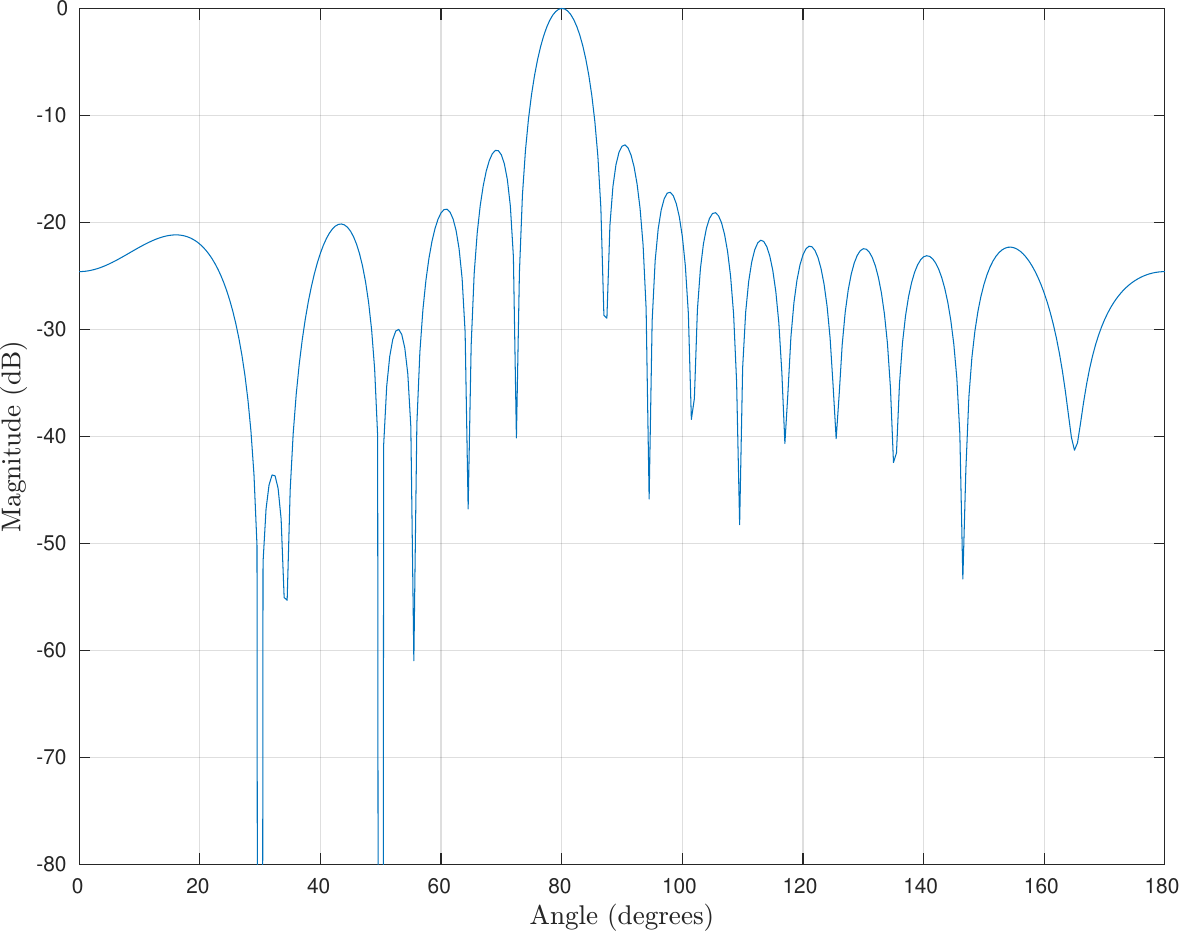}}	
	\subfigure[Nulling with $\theta_c$=$80^o$, $\theta_\text{eav}$=\{$70^o$,$90^o$\}.]{\includegraphics[width=0.329\linewidth]{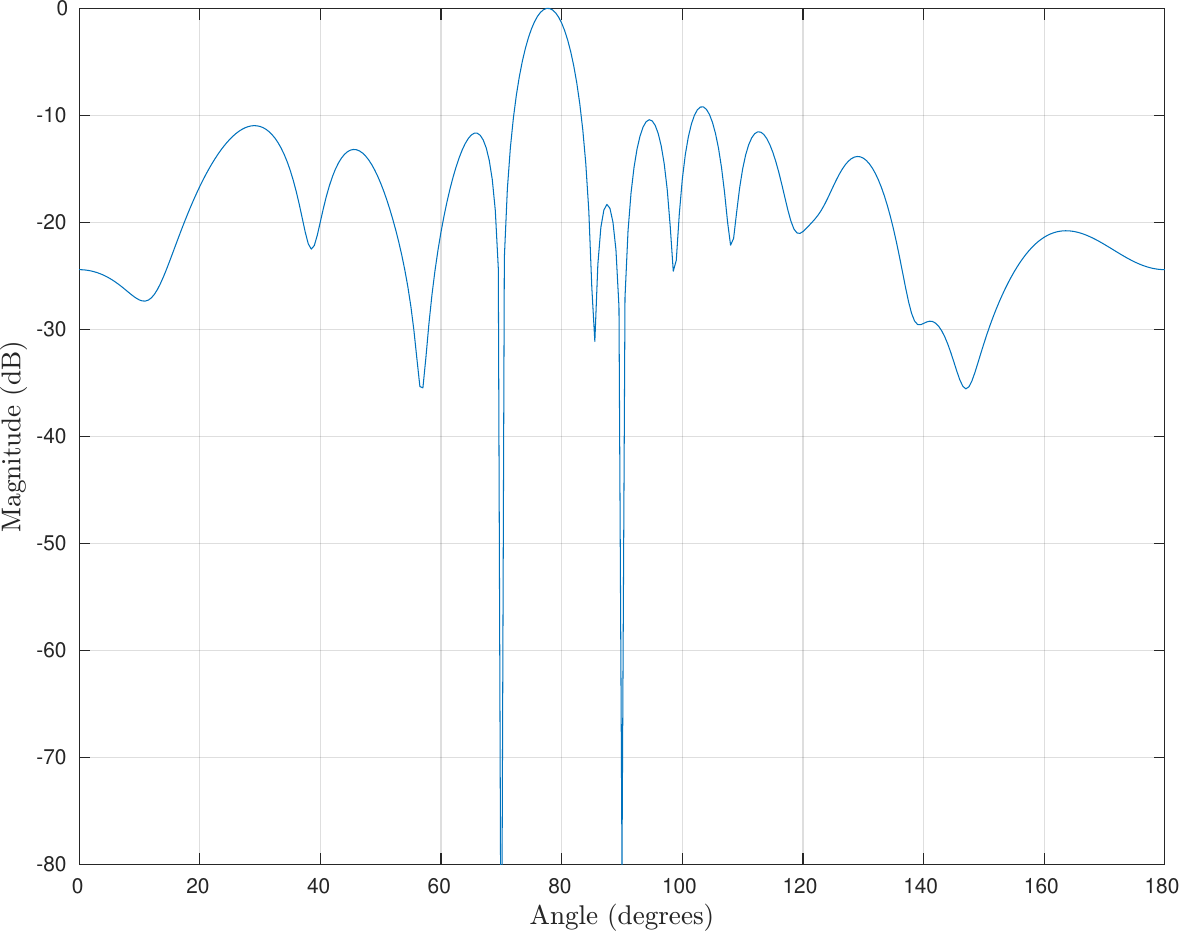}}	
	\subfigure[Proposed with $\theta_c$=$80^o$, $\theta_\text{eav}$=\{$70^o$,$90^o$\}.]{\includegraphics[width=0.329\linewidth]{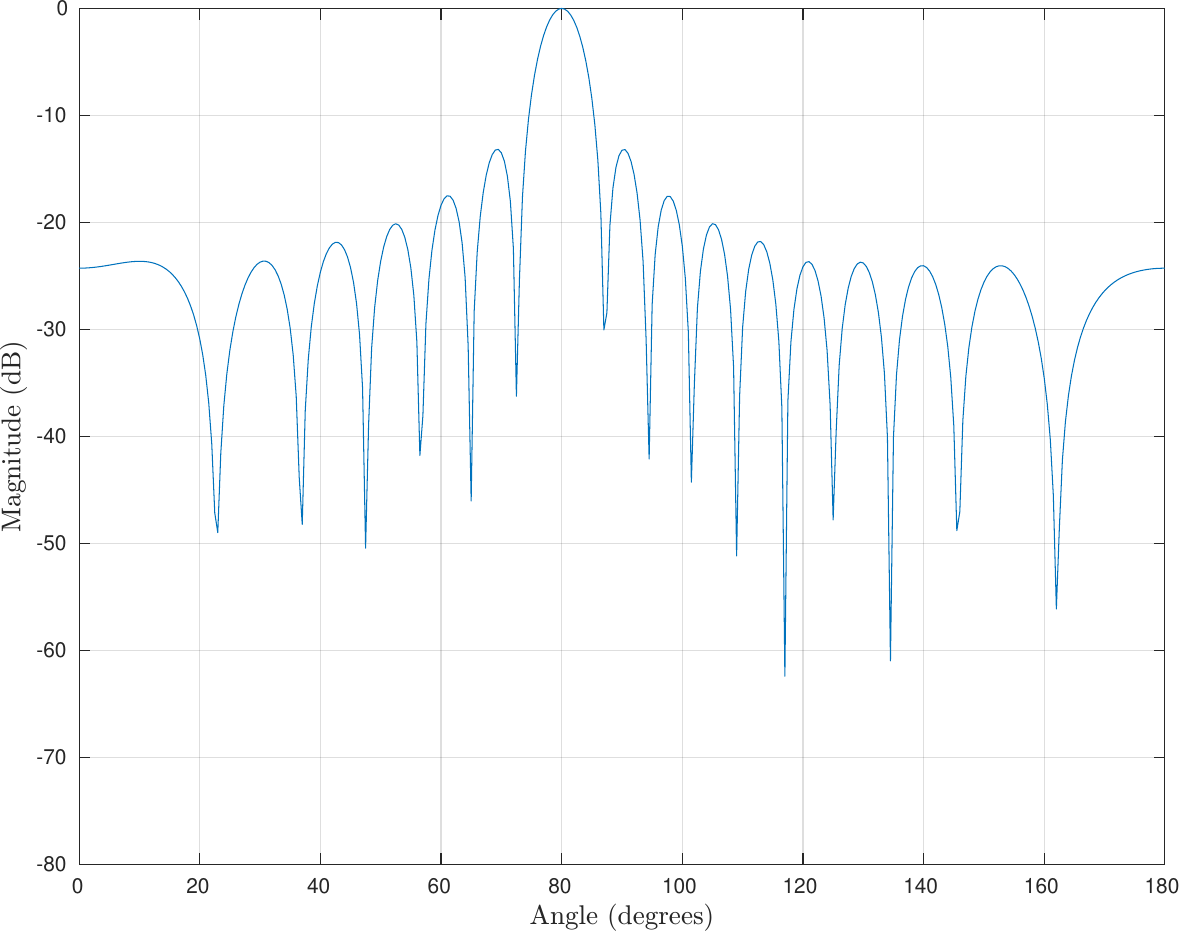}}\caption{Array responses for the nulling techniques in (a), (b), and the proposed DWB (c).} % When the spatial direction of the eavesdroppers is closer to that of the communication receiver in (b), it affects the center of the beam and the peak-to-sidelobe ratio (or dynamic range) which is reduced.}
	\label{fig:array-response}
\end{figure*}

\subsection{Signal Pre-Coding for Eavedropper Deception}
\label{section:precoding}
The challenge now is to find the precise form of the signal we send to the eavesdroppers. This signal must deceive the algorithm that eavesdroppers use for range and Doppler estimation, as was previously described. %First, $\mathbf{D}_\text{e}$ should be an OFDM signal but with the insertion of artificial Doppler and range as described in~\cite{cnf_2023_radarconf1}. 
It does not need to be related to the legitimate information symbols. The best choice in terms of hiding information is to avoid sending the same symbols towards the direction of the eavesdroppers. Hence, we propose that the emitted signal be:
\begin{align}
	\mathbf{d}_{e,i}=\mathbf{F^{H}}\mathbf{\tilde{H}}_\text{sp}\mathbf{x}_{e,i},
\end{align}
where \textit{$\mathbf{\tilde{H}}_\text{sp}$ is the diagonal deceiving channel matrix} and $\mathbf{x}_{e,i}$ can be any random group of QAM symbols destined for the $i$-th eavesdropper. Repeating the derivation in~\eqref{eqn:y-DFT} now for the eavesdropper we have:
\begin{align}
\tilde{\mathbf{y}}_{e,i}=\mathbf{F}\mathbf{y}=\mathbf{F}\mathbf{H}_i\mathbf{F^{H}}\mathbf{\tilde{H}}_\text{sp}\mathbf{x}_{e,i}+\tilde{\mathbf{n}}=\mathbf{\tilde{H}}_i\mathbf{\tilde{H}}_\text{sp}\mathbf{x}_{e,i}+\tilde{\mathbf{n}}.
\label{eqn:y_e_i}
\end{align}
The above expression allows us to understand how $\mathbf{\tilde{H}}_\text{sp}$ should be created: The collective impact of the perceived channel is now $\mathbf{\tilde{H}}_i\mathbf{\tilde{H}}_\text{sp}$ for the $i$-th user. The $\ell$-th entry in the vector $\mathbf{x}_{e,i}$ indicates the symbol that is transmitted in the $\ell$-th subcarrier. If $f_{\ell}$ is the frequency of the $\ell$-th subcarrier we populate this matrix as:
\begin{align}
	[\mathbf{\tilde{H}}_\text{sp}]_{\ell,\ell}= \exp (-j 2 \pi f_{\ell} \frac{R_\text{sp}}{c})\exp (j 2 \pi f_\text{sp} mT_L).
	\label{eqn:H_sp}
\end{align}
In~\eqref{eqn:H_sp} the matrix was written only for the $m$-th symbol. For Doppler deception to work we must use multiple OFDM symbols and apply the same fake Doppler $f_\text{sp}$.  This means that if we run the optimization problem at successive groups of $L$ symbols each, we will change the term $\exp (j 2 \pi f_\text{sp} mT_L)$ according to the OFDM index $m$. This was also seen in~\eqref{eqn:Z_tilde}.

The eavesdroppers will try to estimate Doppler and speed from~\eqref{eqn:Z_tilde}, \eqref{eqn:y_e_i}. It is easy to see that the peak in this fake range-Doppler response will occur at positions $(R_i+R_\text{sp})/c$ and $f_{D_i}+f_\text{sp}$.
Thus, $\mathbf{\tilde{H}}_\text{sp}$ can be created by setting any desired fixed values for the fake range and Doppler ($R_\text{sp}$ and $f_\text{sp}$) without changing them in real-time. The matrix $\mathbf{X}_{e}$, which consists of the individual $\mathbf{x}_{e,i}$ for each eavesdropper, is an optimization variable because the QAM symbols can be any random but valid combination.

\section{Optimization problem \& Solution Algorithm}
\label{section:optimization}
	
\textbf{Optimization Problem for Power Minimization:} For easier formulation of the problem we first vectorize the emitted signal as $\mathbf{s}$=$\text{vec}(\mathbf{S})$. Our beamforming problem targets the minimization of the transmit signal power $\mathbf{s}^\text{H}\mathbf{s}$ subject to the transmission of a deceiving signal to the eavesdroppers and an undistorted signal to the communication receivers. The deceptive wireless beamforming (DWB) problem is:	
	\begin{align}
\underset{\mathbf{s},\mathbf{X}_{e}}{\text{min}}~\mathbf{s}^\text{H}\mathbf{s}\label{eqn:opt-problem1-jcps-pwr2}~~~~
%&\text{s.t. } \mathbf{s}^H\mathbf{s} \leq e_\text{tx},\\
\text{s.t. } &\mathbf{A}(\bm{\theta_\text{e}})\mathbf{S}=\mathbf{F^{H}}\mathbf{\tilde{H}}_\text{sp}\mathbf{X}_{e}\\
&\mathbf{A}(\bm{\theta_\text{c}})\mathbf{S}=\mathbf{D}_\text{c}. \nonumber
\end{align}
The first constraint states that the OFDM signal can consist of any combination of valid QAM symbols $\mathbf{X}_{e}$ that will be spoofed with $\mathbf{\tilde{H}}_\text{sp}$ and fed into the IDFT $\mathbf{F^{H}}$ to produce a deceiving OFDM symbol. The second constraint corresponds to \eqref{eqn:AthetaS}, i.e., an under-determined system of linear equations (when $N_T>N_c$). It is equivalent to a fixed SNR constraint for the communication receivers as in the typical beamforming formulations: The power of the $L$ transmitted samples for the $i$-th receiver is $\|\mathbf{d}_{c,i}\|^2$ which is fixed and it can be scaled by any desired constant. For a given noise power this ratio is equal to the transmit SNR which is controlled by the Tx. Thus, we could actually configure the desired transmit SNR per receiver without having a separate quadratic SNR constraint but just a linear equality. Clearly there are other ways to achieve this, for example, using a direct SINR constraint. An alternative is an approximation in constraint $\mathbf{D}_\text{c}$ (e.g.~\cite{Tang22}), and not the equality, which means that the actual produced signal may differ from the desired one and so it will cause inter-user interference. An advantage of our formulation is that the eavesdroppers are not subject to an SNR/SINR constraint: The optimizer will tend to select a symbol for the $\ell$-th slot of the $i$-th eavesdropper that also minimizes $\mathbf{s}$ and leads to an easier problem. This will lead to lower transmit power and SNR for the eavesdroppers. 

%\TBD{Is it convertible to SDP? Make sure we understand the problem structure. E.g. as in~\cite{Liao11}?}

%-------LONGER
%\textbf{Note:} It is important to note that with the proposed formulation, we do not need to use a performance model or metric for frequency, phase, Doppler, or range estimation at the eavesdropper so that we see how our decision affect the quality of its estimates. This is because even if the eavesdroppers do a perfect estimation, the value that it will be estimated will be fake. 
%-------LONGER
%So focusing on optimizing the transmission through other parameters of interest is enough.

\RestyleAlgo{ruled}
\begin{algorithm}[t]
	$\mathbf{D}_c$ $\leftarrow$ $L N_c$ random QAM symbols\\
	\textbf{Input}~~ $\leftarrow$ $N_T,\text{SNR},L,\mathbf{D}_c$,$\bm{\theta}_c$,$\bm{\theta}_e,\sigma^2$,
	\textbf{Output} $\leftarrow$ $\mathbf{s}^*$\\
	\For{\text{the current OFDM symbol period ($L$ samples)}}{
		$P_s$ $\leftarrow$ $10^\text{SNR}\sigma^2$ -- Calculate power/symbol\\
		$\mathbf{D}_c$ $\leftarrow$ $\sqrt{P_s}\mathbf{D}_c$ -- Set Tx power\\
		Solve~\eqref{eqn:opt-problem1-jcps-pwr2} as QP\\
		round real($\mathbf{x}^\text{opt}_e$) $\rightarrow$ $\mathbf{\hat{x}}^\text{opt}_{e,re}$\\
		round imag($\mathbf{x}^\text{opt}_e$) $\rightarrow$ $\mathbf{\hat{x}}^\text{opt}_{e,im}$\\
		Optimal signal $\mathbf{s}^\text{opt}$ derived
	}
	\caption{Power minimization pseudo-algorithm}\label{alg:cap}
	\label{algo:suboptimal}
\end{algorithm}

\begin{figure}[!htb]
	\centering
	\subfigure[Results for a Tx SNR of \SI{10}{\decibel}, $N_T=16$.]{\includegraphics[width=0.92\linewidth]{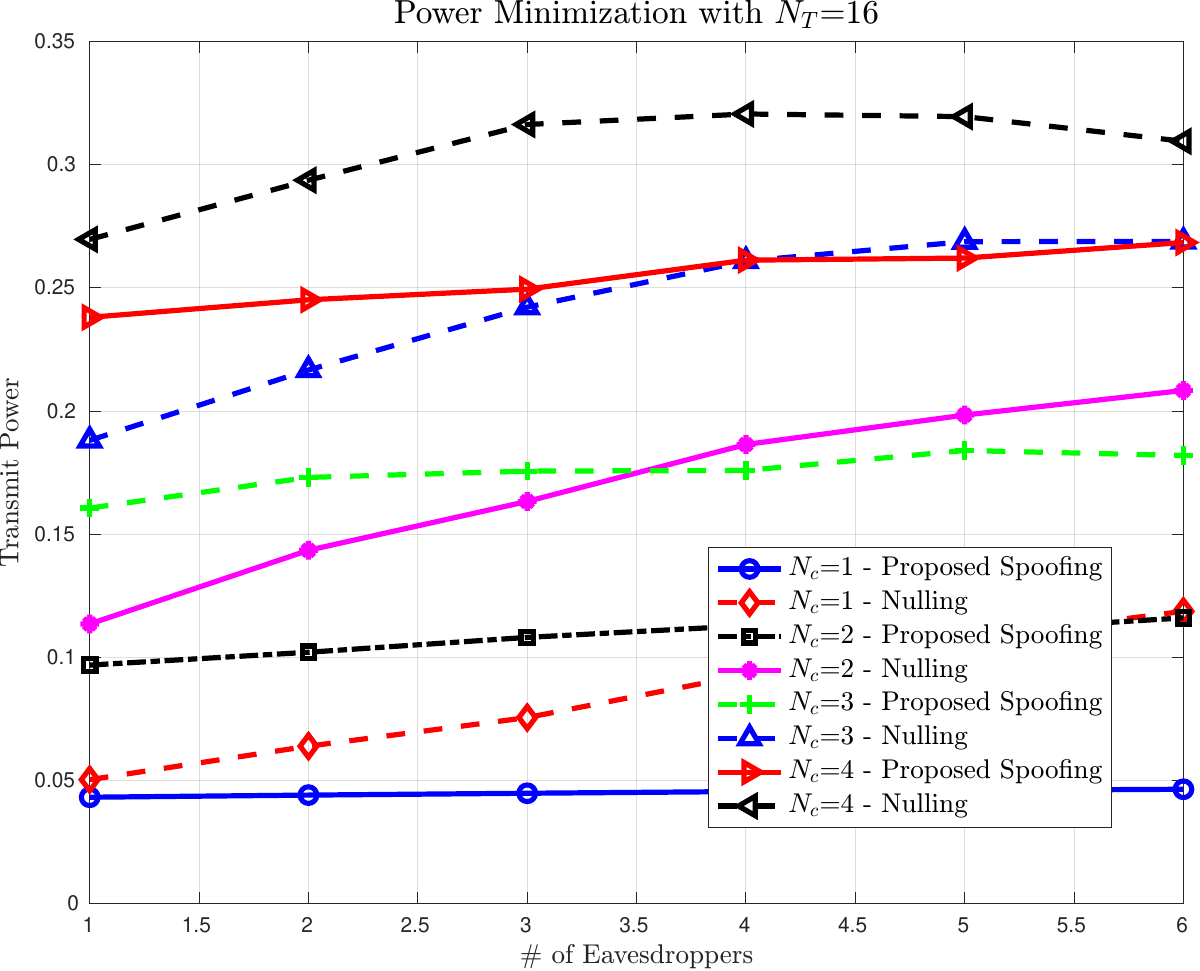}\label{fig:fval-vs-eavesdroppers_ITERS2000}}
%\subfigure{\includegraphics[width=0.99\linewidth]{fval-vs-antennas_ITERS2000}}
	\subfigure[Results for a Tx SNR of \SI{10}{\decibel}, $N_e=4$.]{\includegraphics[width=0.92\linewidth]{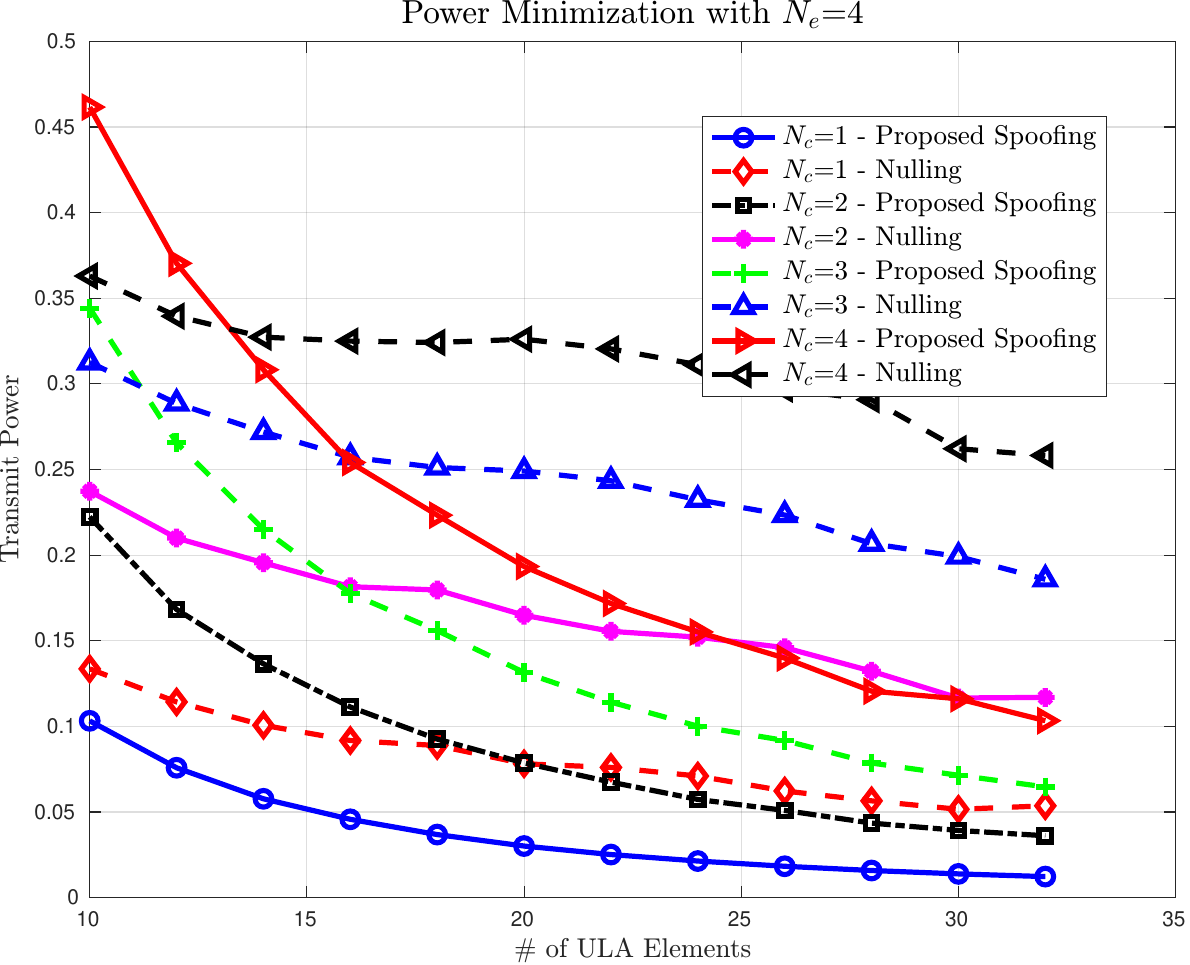}\label{fig:fval-vs-antennas_ITERS3000}}
%\subfigure{\includegraphics[width=0.493\linewidth]{fval-vs-snr_Ne1_ITERS100}}
	\subfigure[Results $N_e=4$ and different $N_c$ and SNR.]{\includegraphics[width=0.92\linewidth]{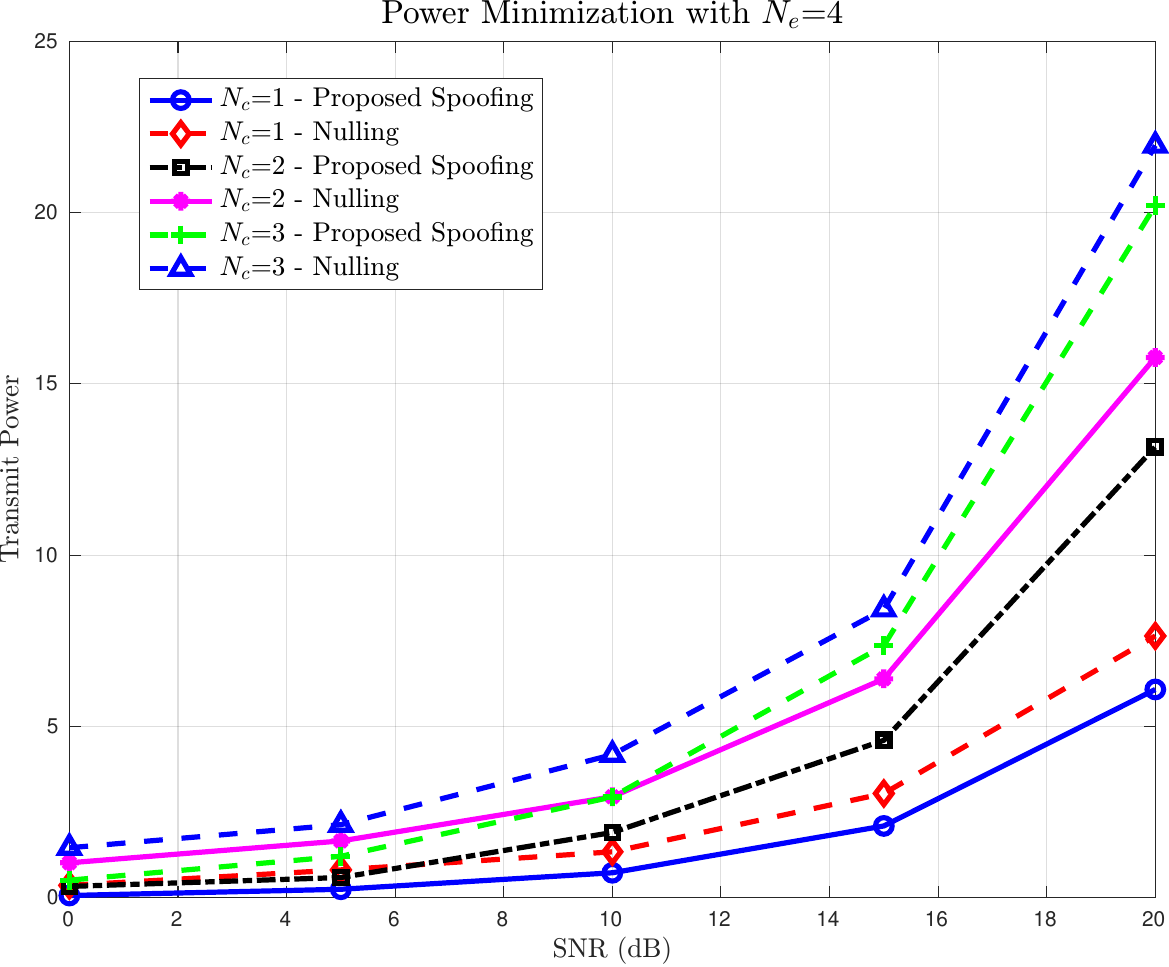}\label{fig:fval-vs-snr}}
	\caption{Simulation results.}
\end{figure}

\textbf{Benchmark Optimization Problem for Power Minimization with Nulling:} If a Tx does not adopt the proposed approach, an alternative way against eavesdropping is nulling. The difference from the problem in~\eqref{eqn:opt-problem1-jcps-pwr2} is that we replace the constraint for the eavesdroppers with $\mathbf{A}(\bm{\theta_\text{e}})\mathbf{S}=\mathbf{0}$, that is we send nothing to them. In this way they ideally receive a signal of zero power and hence we can create a deep null in their spatial direction.

%When we allow continuous variables the problem then is a quadratic problem which is convex and can be solved with interior-point methods.
\textbf{Heuristic Algorithm for the Relaxed Problem:} Unlike the classic beamforming problem for power minimization, in its current form the problem in~\eqref{eqn:opt-problem1-jcps-pwr2} contains additional discrete variables since $\mathbf{X}_{e}$ consists of QAM symbols. As a result,~\eqref{eqn:opt-problem1-jcps-pwr2} is non-convex. If we relax this problem and allow each complex symbols in $\mathbf{X}_{e}$ to take continuous values, then the problem is transformed into a convex QP (QP), i.e., we have a quadratic objective function plus two equality constraints. 
%To solve the relaxed problem with a convex solver we vectorize $\mathbf{x}_e$=$\text{vec}(\mathbf{X}_e)$ resulting in a combined optimization vector $\mathbf{v}$=$[\mathbf{s},\mathbf{x}_e]$. This means that the objective and last constraint in~\eqref{eqn:opt-problem1-jcps-pwr2} takes the form $\mathbf{v^H}\begin{bmatrix}
%	\mathbf{I} & \mathbf{0}\\
%	\mathbf{0} & \mathbf{0}
%\end{bmatrix}\mathbf{v}$, where the matrix in the middle is positive semi-definite. 
With relaxation we allow the symbols transmitted to the eavesdropper to be in the range of the QAM constellation, i.e. $\text{Re}(x_e),\text{Im}(x_e) \in [-\sqrt{P_s},\sqrt{P_s}]$ where $P_s$ is the maximum power per symbol.  
Solving the resulting convex QP and mapping the result to valid QAM symbols is handled by Algorithm~\ref{algo:suboptimal}. The input to the algorithm is an $N_c \times L$ matrix of randomly generated QAM symbols $\mathbf{D}_c$ with energy normalized to unity. Given a desired transmit SNR and a receiver noise power $\sigma^2$ the power of a transmitted symbol is set to $P_s$ Watts. After the QP is solved the optimal solution in the continuous domain is the set of complex vectors $\mathbf{s}^\text{opt},\mathbf{x}^\text{opt}_e$. The valid QAM symbols are those with the minimum Euclidian distance within the chosen QAM constellation. In any case for the transmitter only transmitted signal $\mathbf{S}^\text{opt}$ matters.

%----------LONGER---------------
%\textbf{Note:} An alternative way to solve the problem is to set the quadratic matrix to become $\begin{bmatrix}
%	\mathbf{I} & \mathbf{0}\\
%	\mathbf{0} & \sqrt{P_s}\mathbf{I}
%\end{bmatrix}$. So with this trick we can convert the problem to positive definite QCQP. The objective alters slightly in the sense that we now try also to select symbols for the eavesdropper that have the lowest power (quantity $\mathbf{x_e^Hx_e}$) from the QAM constellation. Still this does not invalidate the idea of the proposed problem formulation.

%\TBD{We use a sub-gradient method to evaluate the globally optimal solution...}.

\section{Simulation Results}
\label{section:performance-evaluation}
\textbf{Setup:} We perform a comprehensive simulation-based evaluation of our scheme. We consider the Tx and various combinations of wireless receivers (eavedropping and communication) that were uniformly distributed at random spatial directions in the range [0,180] degrees, and random distances in the range of [1,100] meters.
We present average power consumption results at the Tx for 1000 random topology configurations using 64-QAM. We explored power consumption versus $N_c$, $N_e$, $N_T$, and the transmit SNR $\|\mathbf{S}^T\mathbf{a}_i\|^2/\sigma^2$ towards the communication receivers. 
%
%-------LONGER
%We also explored different QAM alphabets in terms of the maximum number of symbols they contained. However, we noticed that the QAM alphabet only had impact on performance if higher power was used. So instead of considering different constellations we selected 64-QAM and we explored this aspect in terms of the transmit SNR.

\textbf{Array Response:} To generate the array response we assume the optimal solution is $\mathbf{S}^\text{opt}$. We take the hermitian transpose of its first column to obtain the array response:
\[
B(\theta)=\mathbf{S}^\text{opt,H}[1,:]\mathbf{a}(\theta),~\theta\in [-\pi,\pi]
\]
The steering vector $\mathbf{a}(\theta)$ as a function of $\theta$ was defined in~\eqref{eqn:AthetaS2}. We plot $|B(\theta)|$ for the two systems in Fig.~\ref{fig:array-response}. Fig.~\ref{fig:array-response}(a) depicts the nulling system when the eavesdroppers are farther away from the communication receiver, which is located at $80^{\circ}$. In this case, nulling does not pose a challenge for the optimizer. However, as observed in Fig.~\ref{fig:array-response}(b), when the eavesdroppers are closer in terms of bearing to $80^{\circ}$ nulling increases the sidelobes. Consequently the peak-to-sidelobe (PSL) and the dynamic range are reduced. Furthermore the maximum of the response does not align exactly with $\theta_c$=$80^o$. This is the well-known impact of \textit{nulling} that affects the complete shape of the response. This is not seen in our scheme as the results in Fig.~\ref{fig:array-response}(c) illustrate. By not imposing a nulling constraint on the eavesdroppers we do not affect the PSL ratio, the center of   the response, and in general its shape.

\textbf{Power Consumption:} Results for the power minimization problem in~\eqref{eqn:opt-problem1-jcps-pwr2} are presented in Fig.~\ref{fig:fval-vs-eavesdroppers_ITERS2000}. We compare the proposed DWB optimization with a system that does \textit{nulling} towards the eavesdroppers. The proposed system uses 20\% to 120\% less power for different configurations of the number of users $N_c,N_e$. An even better performance trend is observed as the number of ULA elements $N_T$ increases in Fig.~\ref{fig:fval-vs-antennas_ITERS3000}. As seen earlier our system can create an array response with higher PSL towards the communication receivers while allowing greater flexibility in adjusting power towards eavesdroppers as the first constraint in~\eqref{eqn:opt-problem1-jcps-pwr2} indicates. On the contrary the nulling system has a proportional power increase for higher $N_c$.
Results comparing power consumption versus transmit SNR for two settings of $N_e,N_c$ are plotted in Fig.~\ref{fig:fval-vs-snr}. As expected the transmitted power increases as the transmit SNR increases. However, for higher SNR values, the performance gap starts to widen.

%\antonis{Some ideas for additional results in~\cite{Lagunas20}.}

\section{Conclusions}
\label{section:conclusions}
In this paper we proposed the concept of deceptive wireless beamforming (DWB) to tackle eavesdroppers in OFDM wireless communication systems. DWB deceives the eavesdroppers in terms of range and Doppler (velocity) instead of nulling the signal towards them. The beamformer is designed by solving a relaxed QP. Simulation results indicate that DWB can lead to a beamformer design that achieves very low transmission power, preserves the beam shape, and ensures the privacy of two location parameters of the Tx.
	
	\vspace{-0.2cm}
	
	\bibliographystyle{IEEEtran}
\bibliography{../../../../tony-bib,../../../../MyLibrary}

\begin{thebibliography}{10}
\providecommand{\url}[1]{#1}
\csname url@samestyle\endcsname
\providecommand{\newblock}{\relax}
\providecommand{\bibinfo}[2]{#2}
\providecommand{\BIBentrySTDinterwordspacing}{\spaceskip=0pt\relax}
\providecommand{\BIBentryALTinterwordstretchfactor}{4}
\providecommand{\BIBentryALTinterwordspacing}{\spaceskip=\fontdimen2\font plus
\BIBentryALTinterwordstretchfactor\fontdimen3\font minus
  \fontdimen4\font\relax}
\providecommand{\BIBforeignlanguage}[2]{{%
\expandafter\ifx\csname l@#1\endcsname\relax
\typeout{** WARNING: IEEEtran.bst: No hyphenation pattern has been}%
\typeout{** loaded for the language `#1'. Using the pattern for}%
\typeout{** the default language instead.}%
\else
\language=\csname l@#1\endcsname
\fi
#2}}
\providecommand{\BIBdecl}{\relax}
\BIBdecl

\bibitem{book:fundamentals-of-radar-signal-processing}
M.~A. Richards, \emph{Fundamentals of Radar Signal Processing}.\hskip 1em plus
  0.5em minus 0.4em\relax McGraw-Hill Professional, 2005.

\bibitem{Lagunas20}
E.~Lagunas, A.~Pérez-Neira, M.~Lagunas, and M.~Vazquez, ``Transmit beamforming
  design with received-interference power constraints: The zero-forcing
  relaxation,'' in \emph{IEEE International Conference on Acoustics, Speech and
  Signal Processing (ICASSP)}, 2020, pp. 4727--4731.

\bibitem{Liao11}
W.-C. Liao, T.-H. Chang, W.-K. Ma, and C.-Y. Chi, ``{QoS-Based Transmit
  Beamforming in the Presence of Eavesdroppers: An Optimized
  Artificial-Noise-Aided Approach},'' \emph{IEEE Transactions on Signal
  Processing}, vol.~59, no.~3, pp. 1202--1216, 2011.

\bibitem{Ahmed19}
S.~Ahmed and B.~A. Bash, ``Average worst-case secrecy rate maximization via uav
  and base station resource allocation,'' in \emph{2019 57th Annual Allerton
  Conference on Communication, Control, and Computing (Allerton)}, 2019, pp.
  1176--1181.

\bibitem{Daly09}
M.~P. Daly and J.~T. Bernhard, ``{Directional Modulation Technique for Phased
  Arrays},'' \emph{IEEE Transactions on Antennas and Propagation}, vol.~57,
  no.~9, pp. 2633--2640, 2009.

\bibitem{Tang22}
B.~Tang and P.~Stoica, ``{MIMO Multifunction RF Systems: Detection Performance
  and Waveform Design},'' \emph{IEEE Transactions on Signal Processing},
  vol.~70, pp. 4381--4394, 2022.

\bibitem{argyriou2020}
A.~Argyriou, ``Secrecy in wireless communication through closed-loop receiver
  de-synchronization,'' \emph{Physical Communication}, vol.~43, p. 101195, Dec.
  2020.

\bibitem{cnf_2023_radarconf1}
------, ``{Range-Doppler Spoofing in OFDM Signals for Preventing Wireless
  Passive Emitter Tracking},'' in \emph{IEEE Radar Conference (RadarConf23)},
  San Antonio, Texas, 2023.

\bibitem{jnl_2023_access}
------, ``{Obfuscation of Human Micro-Doppler Signatures in Passive Wireless
  RADAR},'' \emph{IEEE Access}, vol.~11, pp. 40\,121--40\,127, 2023.

\bibitem{chrysanidis2024}
G.~Chrysanidis, Y.~Liu, and A.~Argyriou, ``A {{Replay Attack Against ISAC
  Based}} on {{OFDM}},'' \emph{IEEE Access}, vol.~12, pp. 20\,998--21\,003,
  2024.

\bibitem{Berger10}
C.~R. Berger, B.~Demissie, J.~Heckenbach, P.~Willett, and S.~Zhou, ``{Signal
  Processing for Passive Radar Using OFDM Waveforms},'' \emph{IEEE Journal of
  Selected Topics in Signal Processing}, vol.~4, no.~1, pp. 226--238, 2010.

\bibitem{Braun14}
M.~Braun, ``{OFDM Radar Algorithms in Mobile Communication Networks},'' Ph.D.
  Thesis, Karlsruher Institut fur Technologie, 2014.

\end{thebibliography}

\end{document}